\documentclass{article}
\usepackage{graphicx} 

\usepackage[longnamesfirst]{natbib}

\usepackage[font=scriptsize]{caption}

\usepackage[english]{babel}

\usepackage{amsmath}
\usepackage{amsfonts}
\usepackage{amssymb}
\usepackage{mathtools}
\usepackage{dsfont}
\usepackage{accents}
\usepackage[mathscr]{eucal}
\usepackage{braket}
\usepackage{soul}
\usepackage[normalem]{ulem}

\usepackage{titleref}
\usepackage{varioref}

\usepackage{enumitem}

\usepackage[usenames,dvipsnames]{xcolor}

\usepackage{epsfig}
\usepackage{graphicx}

\usepackage{anysize}

\marginsize{3cm}{3cm}{2cm}{2cm}

\title{Relational objectivity in presence of finite quantum resources}

\author{Luis C.\ Barbado and \v{C}aslav Brukner}

\date{\today}

\begin{document}

\maketitle

\begin{abstract}
The no-go theorems of Bell and Kochen and Specker could be interpreted as implying that the notions of system and experimental context are fundamentally inseparable. In this interpretation, statements such as ``spin is 'up' along direction $x$'' are relational statements about the configurations of macroscopic devices which are mediated by the spin and not about any intrinsic properties of the spin. The operational meaning of these statements is provided by the practically infinite resources of macroscopic devices that serve to define the notion of a direction in three-dimensional space. This is the subject of ``textbook quantum mechanics'': The description of quantum systems in relation to an experimental context. Can one go beyond that? Relational quantum mechanics endeavors to provide a relational description between any quantum systems without the necessity of involving macroscopic devices. However, by applying ``textbook quantum mechanics'' in such situations, it implicitly assumes infinite resources, even for simple quantum systems such as spins, which have no capacity to define an experimental context. This leads to conceptual difficulties. As an alternative, we analyse Penrose's spin network proposal as a potential formalisation of quantum theory that goes beyond the textbook framework: A description in presence of finite resources, which is inherently relational and inseparable in the system-context entity.
    
\end{abstract}


\section{The need for the notion of relational objectivity}

All our statements in physics are given in the form of propositions that use terms associated with classical concepts. This is not only a common practice but, according to Bohr, a necessity in order to understand and communicate to others what we have measured in the laboratory \citep[pp.~25-26]{Bohr1958-BOHAPA}:

\begin{quotation}

\textit{``In this respect we must, on the one hand, realize that the aim of every physical experiment – to gain knowledge under reproducible and communicable conditions – leaves us no choice but to use everyday concepts, perhaps refined by the terminology of classical physics, not only in all accounts of the construction and manipulation of the measuring instruments but also in the description of the actual experimental results \ldots''}
    
\end{quotation}

To illustrate this let us consider the specific case of the spin projection measurement of a spin one-half particle. A rather complicated experimental technique, in which a source of spin particles is prepared, the particles are passed through a Stern-Gerlach apparatus with the magnet aligned along, say the direction $\Vec{n}$ (where $\Vec{n}$ is the unit vector in the three-dimensional Euclidean space), and finally detected at one of the two detectors behind the magnet, can be summarised in a single sentence: \textit{``The magnet was aligned along the direction $\vec{n}$ and the upper detector clicked''}. As such this sentence is a purely descriptive statement of what the experimental arrangement was and what was observed in a single run of the experiment. However, can we conclude from it that ``spin-up along the direction $\vec{n}$'' is an objective property of the quantum system that exists prior to and independent of the experimental context? 


According to textbook quantum mechanics, we assign a quantum state, formalized as a vector in a two-dimensional complex Hilbert space, to the spin particle. The state encodes the probabilities for the outcomes of all possible measurements that can be performed on the particle. This may suggest that statements like ``the spin is up along the direction $\vec{n}$'' refer to the particle {\it alone} as its objective property, independent of whether and which experiment is actually performed. However, from an operational perspective, the statement ``The spin is up along the direction $\vec{n}$'' refers to the entire experimental context and is meaningful only if we have a way to account for $\vec{n}$ as a well-defined orientation in a three-dimensional space. In the laboratory, this is achieved by utilizing \textit{classical} physical devices that can serve as \textit{practically infinite resources} of directional degrees of freedom, such as a macroscopic pointing arrow. Also, to complete the measurement and read out the result, we need a macroscopic device that determines whether the result was ``aligned'' or ``anti-aligned'' with the direction $\vec{n}$, such as a particle detector that ``clicks''.

Consider now a situation where the particle's spin is found to be up along the direction $\vec{n}$ after the measurement. If the measurement was non-destructive for the particle, the detection event can be used as a preparation for a subsequent measurement. If we want to measure the spin along another direction, say $\vec{m}$, it would be necessary to use another macroscopic pointing arrow. Quantum mechanical calculus predicts that the probability $p$ that the spin is aligned along the direction $\vec{m}$ is determined by the angle $\theta = \measuredangle{(\vec{n},\vec{m})}$ between the first $\vec{n}$ and the second direction $\vec{m}$ of the macroscopic pointing arrows, as $p=\cos^2 (\theta/2)$. We observe that all the statements and predictions that we were making, which initially appeared to refer to the quantum system alone, are ultimately about the relative configurations of the macroscopic devices as mediated by the (microscopic) spin one-half system. Consequently, we conclude that the formal quantum mechanical description of the system is \textit{not}, in fact, describing any intrinsic properties of the spin in isolation. Rather, it is describing its properties in relation to the macroscopic devices that define the experimental context.

It can be argued that any physical arrow necessitates only a finite, albeit large, amount of resources to be specified. Therefore, the larger the set of values between $0$ and $\pi$ for the relative angles to be prepared between two arrows, the greater the amount of resources required to be specified. However, as for any macroscopic reference frame these resources exceed those of any finite-dimensional quantum system by far, we will consider the idealised limit of ``practically infinite resources'', where classical relational configurations (such a continuum of possible angles between physical arrows) emerge. The preceding discussions demonstrate that textbook quantum mechanics is inherently reliant on the use of these macroscopic  reference frames in order to adequately describe the preparation and measurement of single quantum systems. In Chapter 3, we will introduce Penrose's spin network proposal as a potential formalisation of quantum theory that takes into account the finiteness of resources.

The necessity of introducing further physical systems, like the macroscopic arrows in the quantum theory of spins, in order to describe the systems of interest is not a novel concept in physics. This is also evident in classical mechanics, where the concept of \textit{reference frames} plays an important role. These are also essential for providing physical meaning to, for instance, the orientation of the rotation axis of a gyroscope. Nevertheless, in the context of classical mechanics and classical statistical physics, this necessity, although present, can be ignored without compromising the physical description. This is because in these theories, one can assign a joint probability for the values of all possible measurement records for all possible measurements on the classical system. For example, any macroscopic reference frame for spatial directions allows the assignment of a joint probability to all components of the gyroscope's axis of rotation. As a result, the choice of macroscopic reference frame for preparing or measuring the gyroscope is arbitrary. Consequently, the identification of the components of the axis of rotation as intrinsic properties of the gyroscope within the formalism does not conflict with its ultimately relational nature.

The situation is fundamentally different in quantum mechanics. The no-go theorems of Bell \citep{PhysicsPhysiqueFizika.1.195}, as well as those of Kochen and Specker \citep{Koc}, show that in general one cannot assign truth values, not even probabilistically, to the statements about measurement results for
(actual as well as counterfactual) measurements prior to and independent of the experimental context---the entire classically describable measurement arrangement. 
There are basically two ways to respond to the no-go theorems. One can still decide in favour of retaining the classical notion of objectivity, in which one ascribes well-defined properties to a system, but which change when the experimental context changes. Bohmian mechanics is an interpretation of quantum theory that follows this view \citep{Bohm1952-BOHASI}.

An alternative view is based on the understanding that at the core of the theorems of Bell, Kochen, and Specker is \textit{an a priori assumption of separability between the system and the observational context.} It is only when this assumption is fulfilled that it is at all conceivable to define what the system is on its own. One can deny this separability and take the view that the context is a constitutive element that, jointly with the system, defines in relational manner what is objective in quantum theory\footnote{While we will not discuss this further here, in our opinion, this does not require the presence of an agent or a laboratory; the experimental context could be defined by the prevailing type of decoherence in the environment of the system far from any agents or laboratories. However, even in such situations it might be useful to imagine fictitious agents that can actually use quantum theory to predict probabilities for individual results in a given experimental context or even choose different contexts.}. This is the core idea behind the notion of \emph{relational objectivity:} What is objective in quantum mechanics are not the intrinsic properties of the quantum system, but its specific relation with a \emph{concrete physical macroscopic} context. In this sense, the notion of a closed quantum system in the ``textbook quantum formalism'' would only be a shorthand notation for describing the physical relations between the given quantum system and the observational context. Therefore, the ``formal-theoretically closed'' quantum system would not describe an ``isolated system'' but its relations to macroscopic devices. And while, as discussed previously, in the classical case this lack of isolation has no further consequences, in the quantum case it is conceptually critical. For example, a spin one-half particle is then never objectively oriented ``up'' along the direction $\vec{n}$, but is objectively oriented along \emph{this macroscopic red wood arrow here in the lab.} Given any other macroscopic blue plastic arrow, the spin orientation is \emph{undefined along the blue plastic arrow.} The probability that it will be aligned or anti-aligned along the blue arrow, after projection onto it, given that it was previously aligned along the red arrow, is given by an objective property \emph{not of the spin particle, but of the macroscopic relation between the two arrows}, defined by the relative angle between their two directions. In this way, the absence of a joint probability distribution for all possible measurements on the spin is only natural, since the formalism never provides the probabilities for intrinsic properties of the spin, but rather for the results of a given \emph{specific} experimental context. Any counterfactual reasoning about what ``unmeasured properties would have been if another measurement had been made'' makes no sense if it is fundamentally impossible to separate the notions of system and context. Interpretations of quantum theory that are in agreement with this view include Quantum Bayesianism (QBism) \citep{PhysRevA.65.022305}, the relational \citep{rovelli1996relational, Rovelli2005, sep-qm-relational}, Healey's pragmatist approach \citep{Healey2012-HEAQTA}, as well as various interpretations that are either inspired by or are modern expressions of the Copenhagen interpretation \citep{bub2017bohrmostlyright,BUB2020251,Brukner2017,landsman2017foundations,janas2021understanding,healey_revolution,pittphilsci22476,auffeves2016contexts}.

In summary, we have exposed two critical features of quantum mechanics: The relational nature of any quantum state with respect to an experimental context, \emph{and} the finiteness of the resources of any physical system, including the experimental context, with the textbook quantum mechanical description being recovered in the limit of infinite resources of the context. Is there a description of quantum systems that incorporates, by construction, these two features? We approach Relational Quantum Mechanics as a natural candidate, and conclude that, while it embraces a relational description (as the very name indicates), it fails in taking into account the finiteness of the resources, which leads to several conceptual problems. We then explore the original construction of spin networks by Roger Penrose, which naturally incorporates \emph{both} desired features. This leads us to conclude that such construction is more adequate for the understanding of contextuality and non-locality in quantum mechanics, as well as the potential extension of the quantum framework beyond the conventional textbook quantum mechanics.




\section{`Relational Quantum Mechanics' and the problem of insufficient resources}

Relational Quantum Mechanics (RQM) is an interpretation of quantum mechanics that takes a relational view of the properties of quantum systems as its fundamental premise \citep{rovelli1996relational, Rovelli2005, sep-qm-relational}. For RQM, properties of quantum systems are only meaningful in relation to other systems. In this sense, it is perfectly compatible with the theorems of Bell and Kocher and Specker.

In its attempt to extend the scope of the ``textbook quantum theory'' to the domain outside the laboratory paradigm, where the notions of preparation, measurement, observation, context, and observer apparently play no role, RQM proposes relational objectivity between \textit{any} two quantum systems, not only between a quantum system and a classically describable experimental setup. According to RQM, any interaction between two quantum systems should be considered as a kind of ``quantum measurement'' in which the systems reveal to each other some \emph{facts,} these facts being meaningful only in relation to the other system. In this manner, RQM also aims to resolve the ambiguity of the quantum measurement problem inherent to the laboratory paradigm. This ambiguity pertains to the question of which devices in the laboratory induce unitary evolution and which projections as a result of a measurement. In RQM, it is asserted that there is no ambiguity, as both types of evolution represent an interaction between quantum systems, and there is no fundamental distinction between the two. 

Recently, however, several problematic aspects of the RQM program have been raised in the form of no-go theorems \citep{pienaar2021quintet,brukner2021qubits,mucino2022assessing}. In particular, the theorem in \citep{brukner2021qubits} demonstrates that RQM's notions of ``knowledge'', ``information'', or ``facts'' that one quantum system may assign to or have in relation to another are incompatible with the fundamental features of the notions that are common in everyday life and physics praxis: (1) \textit{the ``observer knows'' the state of the ``observed system'' when the states of the two are correlated,} (2) \textit{distinguishable ``facts'' are represented by orthogonal quantum states}. 

We illustrate the essence of the no-go theorem with an example. Imagine that, from the point of view of C, two spin one-half particles A and B are prepared into the singlet state -- the totally anti-symmetric Bell state. 
According to RQM, the correlations between the two spins give rise to facts that A acquires relative to B and vice versa, of form ``spin B is `up' (`down') along direction $z$ relative to spin A''. This is in agreement with condition (1). However, since \textit{no basis is preferred for two spins}, one can rewrite the singlet state in an arbitrary but same basis of the two spins, and its form remains invariant. This leads to infinitely many further facts of form ``Spin B is `up' (`down') along \textit{arbitrary} direction $\vec{n}$ relative to spin A''. However, distinguishable facts like ``orientation `up' along direction $z$'' and ``orientation `up' along direction $x$'', are \textit{not} represented by orthogonal states, not even approximately, as long as they are represented by sufficiently low-dimensional systems such as spin one-half particles. This contradicts (2).

The two fundamental features (1) and (2) of the notion ``knowledge'' align closely with Bohr's quote from the beginning of the chapter. In the laboratory paradigm, the potential ``facts'' that an observer can ascertain about a quantum system or that can be stored in a measurement  device are distinguishable and thus represented by orthogonal (or nearly orthogonal) quantum states in the preferred basis induced by decoherence. This allows for the information about the facts to be read out unambiguously by multiple observers, communicated, and copied, ultimately leading to intersubjectivity about the facts among the observers. In contrast, the states between two correlated quantum systems are, in general, not orthogonal and cannot be read out unambiguously or copied -- that is, ``qubits are not observers.''

The example provided for the no-go theorem already brings out the main problems in which RQM runs from a conceptual point of view: The central tenet of the RQM interpretation, namely the assumption that any interaction between two quantum systems can be regarded as acquiring certain facts of one quantum system about the other, in the form of outcomes of quantum measurements, is in direct conflict with the limited resources of microscopic quantum systems. These resources simply are insufficient to specify a quantum measurement. For example, in order for a quantum system to be able to measure the orientation of the spin of a spin one-half particle, it should first and foremost be able to determine a direction and orientation in space with sufficient accuracy, which requires a large amount of directional resources. A magnet made up of a large number of equally prepared spin particles produces a magnetic field with a well-defined direction and orientation, which can then be used to determine a spatial direction along which other spin particles can be measured. But, for example, a half-spin particle alone is incapable of such a task because it alone cannot determine a direction in space. Assuming the opposite leads to contradictions as pointed out in \citep{brukner2021qubits}. In addition, according to RQM a quantum system has the ability to act as a measuring device, manifesting ``facts'' relative to this system, for which quantum mechanically probabilistic predictions can be made. However, if we were to consider a simple quantum system that could act as a measuring device, we would face the same kind of questions that are common to the measurement problem: When, under what conditions and in which basis is such a ``fact'' realized? What is the state of the system after the fact is realized? What is the rate of ``probabilistic transitions'' between past and new facts? We see that instead of solving the measurement problem, RMP installs it where it never existed, namely in situations where there are no measurements. 

In their response to critics \citep{di2022relational}, the authors state that ``the information that a qubit has about another qubit need not be such that we can understand what subjective experience of knowledge it would correspond to, and therefore RQM is under no obligation to solve the preferred basis problem''. But the no-go theorems never talked about ``subjective experience'', they do not require the existence of an agent or even a measurement, they simply state that RQM's notions of ``knowledge'', ``information'' or ``facts'' are incompatible with the very properties we use to identify these notions, if they were to be applicable to systems with limited resources. So it is RQM's task to explain what they mean, how we can operationally verify what they are supposed to mean, or to introduce new terms that avoid being misinterpreted as having anything in common with the conventional meaning of the terms.   

Another counterargument given in \citep{di2022relational} is that ``RQM is about facts, not states''. According to this, the argument in \citep{brukner2021qubits} would be mistaking the different relational descriptions of different systems, namely the information on facts about one qubit as revealed to the other qubit, and the information on the fact that the two qubits hold a relation as described by a third system. We will take up the argument in \citep{brukner2021qubits} again, and reformulate it within the postulates of RQM itself \citep{rovelli1996relational}. We will show that the presumed distinction between facts and states is not tenable according to these postulates, and therefore show that the contradiction pointed out is still present and inherent in the RQM program, and that it ultimately has to do with the lack of resources we argued in the previous paragraph.

Consider again the two spin one-half particles A and B interacting with each other. As described by a third observer C, the particles are assumed to get entangled, again in the singlet Bell state; while for the systems themselves, A has revealed some fact about the other particle B, this fact being its spin along some direction. In terms of the postulates of RQM \citep{rovelli1996relational}, the act of interaction may be interpreted as A asking B the question ``Is your spin oriented along the direction $\vec{n}$?'', where $\vec{n}$ is a unit vector in the three-dimensional space, and receiving a ``yes/no'' answer. Note that, according to RQM, the set of questions that can be asked of a system does not depend on the resources the questioner has. This is the root of the problem: RQM tries to reproduce the structure of a quantum mechanical measurement where there are not enough physical resources to do so.

According to Postulate~1 of RQM \citep{rovelli1996relational} ‘There is a maximum amount of relevant information that can be extracted from a system,’ and since in the case of spin one-half this maximum is one bit, according to RQM particle A already exhausted the maximum of relevant information about B. Next, according to Postulate 2, ‘It is always possible to acquire new information about a system', but since A already has the maximal relevant information about B, the probabilities of the answers to any new question A wants to ask B are determined by the previous information in ``the best way'' possible. At this point, we shall ask: What is the nature of this maximal information that A has about B, and that allows for determining the probabilities of future answers? Notice critically that this information cannot be just one single ``yes/no''. One bit is meaningless without specifying the question to which the one-bit answer is given. The relevant information is that the question ``Is your spin oriented along the direction $\vec{n}$?'' has been answered by ``yes/no''. So the total information consists of a pair, (yes/no, $\vec{n}$), the yes/no bit \emph{and} the three dimensional unit vector $\vec{n}$. Since having a spin aligned along the direction $\vec{n}$ is equivalent to having a spin anti-aligned along the direction $-\vec{n}$, formally $(\textrm{yes}, \vec{n}) \equiv (\textrm{no}, -\vec{n})$, the complete information about the spatial direction and the answer yes/no can be encoded in a vector of the Bloch sphere. Therefore, the space containing this information \textit{must} be (isomorphic to) a projective two-dimensional complex Hilbert space. We conclude that the information on facts about B that A has must be represented by a quantum state\footnote{As more complex finite dimensional quantum systems can be built up from the composition of a finite number of qubits, and this way of encompassing larger systems with the formalism is explicitly embraced by RQM \citep{rovelli1996relational}, the argument provided applies to all the cases for which RQM has been explicitly discussed so far.}.

But, where is this information that A has about B physically stored? What the no-go theorems show is that there is no physical place in A to store the information on the fact/state about B that it has gained during the interaction in a reliable way. We notice that the necessity for information to be encoded in physical degrees of freedom, which is upheld by RQM, plays a pivotal role in the critique. In their response in \citep{di2022relational}, the authors reject the criticisms on the grounds of a naturalistic view of information. However, we believe that their argument is actually put upside down. We quote (changing some dummy labels to maintain consistency with the previous example):

\begin{quotation}
    
\textit{
What is the meaning of the statement that A has knowledge about [...] B?
}

\textit{
There are two possible answers. The first is a naturalistic answer. The second is a dualistic or idealistic answer. According to the first, this is a statement about the actual physical configuration of the ink and the notebooks [...] in A and about the correlation of these with whatever can be observed in B. According to the second, A's knowledge is something over and above its physical configuration. In this case, the ``inaccessibility'' of A's knowledge, namely of the ``universe as seen by A'' is indeed there. But this only follows because one assumes that knowledge is unphysical.
}

\textit{
We adhere to a naturalistic philosophy. In a naturalistic philosophy, what A ``knows'' regards physical variables in A. And \emph{this} is accessible to C. If knowledge is physical, it is accessible by other systems via physical interactions.
}

\end{quotation}

We concur with the characterization of the knowledge according to the naturalistic and idealistic philosophies, and we acknowledge the choice for the naturalistic approach. However, we have demonstrated with a simple example that RQM, in its concrete attempt to reconstruct quantum mechanics, inadvertently falls into a picture closer to the idealistic philosophy. In particular, from what the example shows, it cannot be true that ``what A knows regards physical variables in A.'' This is not possible as there is \textit{no physical resources} in A where the physical information $\vec{n}$, which is the fact/state that A gets to know about B, can be stored in a reliable manner that allows it to be used to determine the probabilities for subsequent facts about in B in a quantum mechanical manner. There are no ink or notebooks, as A is a simple one-half spin particle. In even more straightforward terms, it is incorrect to assume that a one-half spin particle can perform measurements on other spin particles 
as this task is beyond the capabilities of such a simple system.

Therefore, in our view RQM's attempts to evade statements about states as unphysical are merely rhetorical: The physical information in RQM looks and behaves like an ordinary quantum state. We notice that this conclusion was to be expected if RQM aimed to reproduce quantum mechanics in the limit of macroscopic preparation and measurement devices. There, quantum states are always conditioned on past facts about the preparation procedure. In this sense, for example, ``the spin-up along the direction $\vec{n}$'' is a mathematical formalization of the physical fact that the knob on the preparation device was oriented to produce particles with spin-up along the direction $\vec{n}$. Moreover, the quantum state represents also the information needed to compute the probabilities for the responses to the new question ``Is your spin oriented along the direction $\vec{m}$?'' in the ``best way'' possible, as Postulate 2 of RQM requires. Actually, this relation between facts and their representation in terms of quantum states was an integral part in \citep{brukner2021qubits}, and so the counter-argument simply did not apply in the first place. It may only apply if under ``facts'' something else is meant, which is what the no-go theorems prove. Alternatively, one could maintain the conventional meaning of ``facts''; however, this would result in contradictions of RQM with its own postulates.

To conclude this section, let us address the revised interpretation of RQM in \citep{adlam2022information}, where the absence of a preferred basis in correlations between quantum systems is not regarded as a defect but rather as a feature of RQM. It is proposed that
\begin{quotation}
\textit{``The alternative is to agree that quantum events do not typically have the simple form
‘variable $V$ taking value $v$ relative to Alice.’ Rather they must have a conjunctive form: ‘variable
$V_1$ taking value $v_1$ relative to Alice, and variable $V_2$ taking value $v_2$ relative to Alice, ....’ and so
on, specifying definite values for each of the variables [...] 
in all of the different possible bases for it.''}
\end{quotation}
The question arises as to what the physical content and operational meaning of the statement that ``variable $V_i$ takes the value $v_i$ relative to Alice'' for all possible bases $i$ truly is. In the case of two spin one-half particles A and B in the singlet state, these statements have an explicit form of ``spin B along direction $x$ takes value `up' (`down') relative to A'', ``spin B along direction $y$ takes value `up' (`down') relative to A'', etc. In order for these statements to have some explanatory or at least descriptive power, we require that they have a testable consequence when measurements are made, even though we acknowledge that the statements could be assigned to the systems without the necessity of making the measurement. A natural way to establish a correspondence between the abstract statements and observations is to give to the statement ``spin B along direction $x$ takes value `up' relative to A'' an operational meaning that ``whenever B is measured to be `up' along $x$ by a third system C, then A is found to be `down' along $x$ by this same system C, if measured right after'', in the spirit of the cross-perspective links proposed in \citep{adlam2022information}. However, this step brings us dangerously close to the hidden variable trap of the Einstein-Podolsky-Rosen (EPR) program \citep{PhysRev.47.777}. For if one would require that the statements are ``real'' for all directions $\vec{n}$ in the sense that their truth values are independent of whether and which measurement is performed on a distant system B, then there are ``elements of reality'' in A in the EPR sense, and one is trapped in the hidden variable hole. One can try to argue against this by negating the concrete operational correspondence or the reality of truth values. As one can deduce from the whole discussion, we would not know how to fix the first option without leaving the pivotal notions of `fact' or `quantum event' as empty terms with undefined observable consequences, and the second option is essentially the Copenhagen interpretation that RQM wanted to overcome. 

\section{Spin networks as a relational description with finite resources}\label{sec:networks}

The spin networks, in their original formulation by Penrose \citep{penrose2010roger,Penrose1971AngularMA,Penrose1972OnTN}, constitute a framework in which the quantum theory of spins and the structure of the three-dimensional space of directions would be expected to arise together out of more primitive combinatorial rules. Since the aim is to reconstruct space rather than assume it, the combinatorial rules are based on the addition and subtraction of total angular momentum, which is rotationally invariant, does not single out any preferred spatial direction, and therefore does not require a pre-existing space.

Let us briefly review the basics of spin networks, in order to show how they could account for the notion of relational objectivity in standard quantum mechanics as emergent out of a more primitive notion of objectivity. A spin network consists of a graph which line segments, called units, are identified with a non-negative integer each, which represents its total angular momentum in units of $\hbar/2$. The units come together in vertices, exactly three in each vertex. We may think of a vertex as representing the combining of two units together to make a third, or as the splitting of a single unit into two separate units. These combinations shall satisfy the rules for addition of quantum angular momentum, namely the triangular inequality and the conservation of fermions.

\begin{figure}[h]
\begin{center}
\includegraphics[width=0.54\linewidth]{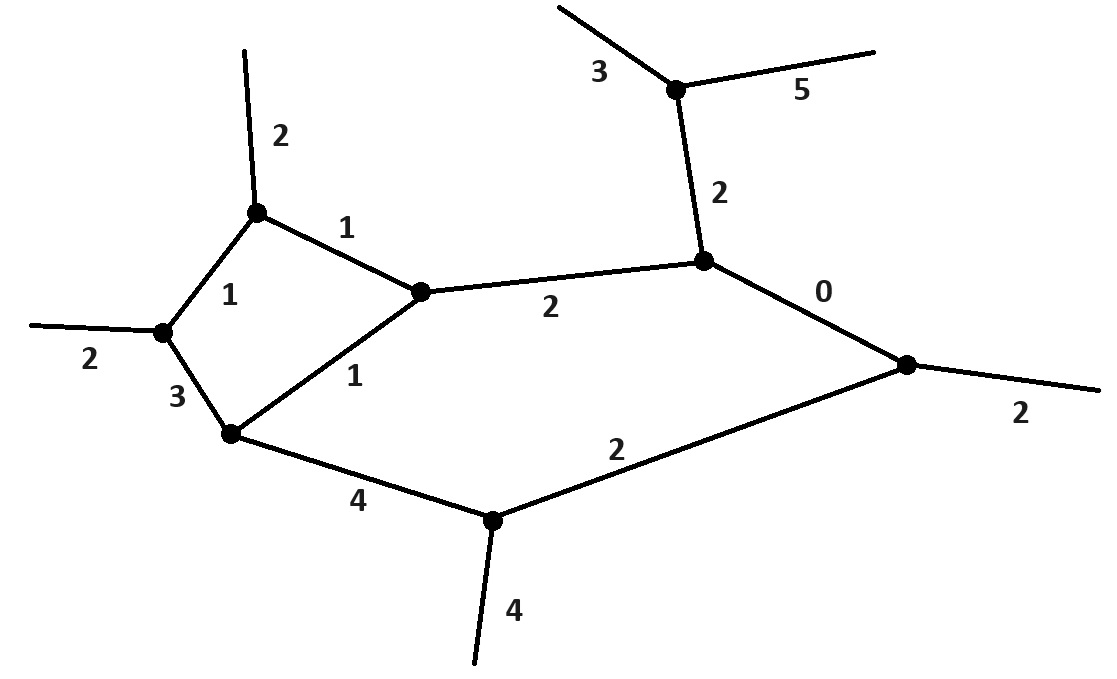}
\caption{An example of a spin network. Line segments are called units, and are labelled with a non-negative integer that represents its total angular momentum in units of $\hbar/2$. Units meet in vertices, three in each, in a way that satisfies the rules of recoupling of quantum angular momentum.}
\label{network}
\end{center}
\vspace{-0.5cm}
\end{figure}

We provide an example of a spin network in Fig.~\ref{network}. The units in the network that are not connected to two vertices are called ``free ends''. Given one network, one can consider the same network where two of the free ends merge together in a new vertex and form a new unit. The rules of spin networks provide a way to compute, through purely combinatorial means, the probabilities of obtaining the different possible values of angular momentum of this new unit \citep{penrose2010roger}. 

The spin geometry theorem \citep{penrose2010roger,moussouris1983quantum} shows that free ends with a large angular momentum can be used to define directions in three-dimensional space. These directions are however of an intrinsic \emph{relational} nature, since what one can define is only the relative angle subtended by two free ends. So spin networks ``build up'' themselves an emergent notion of space in the limit of large spins, rather than living in a pre-existing space.

\begin{figure}[h]
\begin{center}
\includegraphics[width=0.5\linewidth]{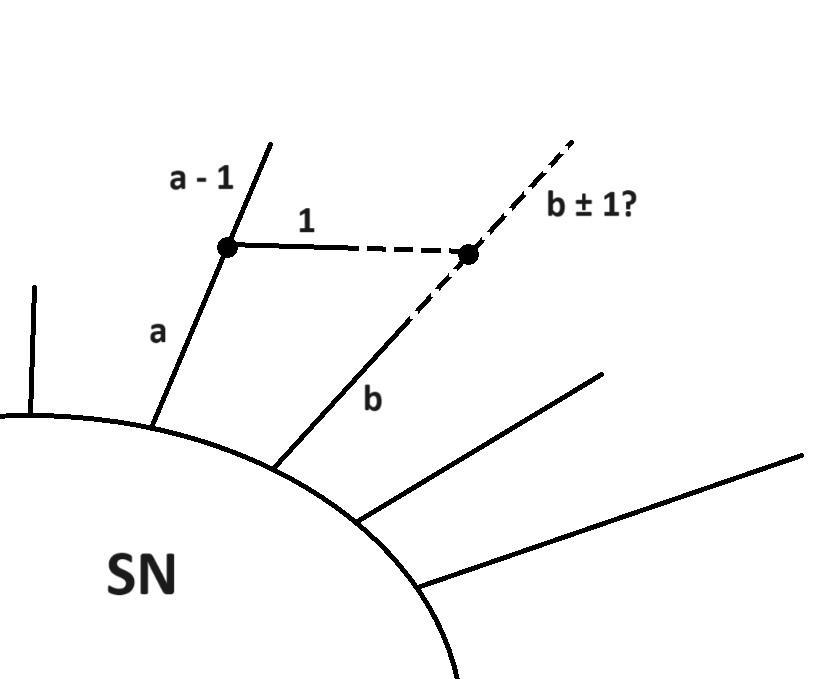}
\caption{An example of a splitting of a one-half spin particle from an end unit $a$, and its later coupling with an end unit $b$. The area denoted as SN stands for a generic spin network. The rules of spin networks allow to compute the probabilities that the value of the new unit formed is $b\pm 1$, given SN.}
\label{coupling}
\end{center}
\vspace{-0.5cm}
\end{figure}

The notion of relative angle between large free ends can be defined in the following way. Consider two free ends, with large values~$a$ and~$b$. Consider a splitting of the first free end, where one of the split units is a one-half spin particle (an unit~1), while the other unit has a value of~$a-1$. One can consider then the combination of the split one-half spin particle with the second free end, and use the rules of spin networks to compute the probability~$p$ that the value of the unit resulting of this combination is~$b+1$. We depict the process in Fig.~\ref{coupling}.

An interpretation of this procedure is the following: Since the one-half spin particle, when split out of the original free end, decreased the angular momentum of it, we can say that such particle was ``prepared as aligned'' with the first free end. If, when combined with the second free end, the particle happens to increase the angular momentum of it, we can say that the particle was ``measured to be aligned'' with that other free end. Now, we can imagine for a moment that these free ends corresponded to strictly well-defined directions in space, through gyroscopes with high angular momentum. We know then from quantum mechanics that the probability~$p$ that a particle aligned with the first gyroscope, when measured along the direction of the second, is found out as aligned with it, is given by the angle~$\theta$ between the two angular momentum vectors of the gyroscopes, as $p = \cos^2 (\theta/2)$. 

In the case of spin networks, as we mentioned, the free ends do not correspond to any direction in a pre-existing space. So we do not compute~$p$ out of~$\theta$. We rather start with the probability~$p$, as given by the rules of spin networks, as the fundamental quantity, and \emph{define}~$\theta$ from the relation above. We can this way compute a collection of angles between any pair of free ends in a spin network. The spin-geometry theorem shows that, for spin networks with large end units, and such that the values of these angles remain stable enough under repetitions of the procedure described, this collection of angles satisfies the rules of spherical geometry, that is, of directions in a three dimensional space, with arbitrary accuracy.

The spin-geometry theorem shows how spin networks give rise both to rotational symmetry of the Euclidean three-dimensional space and rotational degrees of freedom of quantum mechanics. From a formal point of view, this could be considered a trivial result, as the combinatorial rules of spin networks are inspired themselves on the standard recoupling theory of quantum angular momentum. But what we want to emphasize here is the conceptual twist that is suggested, considering the coupling mechanism and the probabilities as fundamental, and the space as emergent. Because it is this conceptual twist that can give rise, as an emergent notion in the limit of large units, to the \emph{relational objectivity} that we discussed for quantum mechanics in the previous sections.

We propose to \emph{identify} the elementary spin network, in the limit of very large spins and well-defined angles, with the elementary laboratory procedure where the preparation and measurement devices are oriented at a fixed relative angle; and to \emph{identify} the split and coupling of smaller units between these devices (as the exchange described in the procedure above) as the quantum preparation and measurement processes\footnote{The identification of the recoupling of spins with larger spin systems as a measurement of the spin orientation is already present in previous works \citep{Poulin_2006,Poulin_2007}. However, in these works the identification is done within the full quantum mechanical formalism (the resemblance to spin networks is only mentioned \citep{Poulin_2006}). While in such case the resources required by the preparation and measurement devices may be explicitly present, the resources required for fixing their relative orientations are not, while in the spin-networks formalism they are through the existence of the rest of the spin network (SN in Fig.~\ref{coupling}), which \emph{prepares} the devices to hold a given angle.}. Since, by the very nature of the framework, spin networks provide the probabilities for only \emph{one} experimental context, there is by construction no room for specifying joint probabilities for the outcomes of all possible experimental setups or for counterfactual reasoning, both of which are used to derive the Bell-like theorems. The large enough spin networks constitute the \emph{inseparable object-context entity} that is required in the relational approach.

We notice how the spin networks with finite units define probabilistic processes that may happen without giving a reference to a measurement. However, the operational significance of these probabilities is unclear. Under standard laboratory conditions, any experimental test to verify these probabilities would require a modification of the original finite spin network to include virtually infinite spin units to represent the infinite directional resources of the laboratory. Thus, any such test would only examine the compatibility of the Penrose framework with the textbook quantum mechanics in the limit of infinitely large spin units. Nonetheless, together with the inherently contextual nature of the preparation and measurement processes in the construction, the other two key facts that we want to highlight about it are: a) Unlike in RQM, the amount of possible relational configurations between systems scales with the resources of those. Therefore, the relational configuration between any two quantum systems, under the limitation of low resources, is not given by \emph{any} possible quantum state of a given Hilbert space, but rather by a much more reduced range of possible relational configurations. This avoids the problems of RQM that we stressed in the previous section, which originated from the attempt to implement the structure of quantum measurements in the whole Hilbert space under the presence of finite resources. b) The quantum mechanical formalism for preparations and measurements, which assigns states in a Hilbert space and computes the probabilities out of them, would be recovered in the limit of large resources (the limit where it has been empirically tested), while at the same time the ``measurement devices'', apart from their large size, have no special status compared to other spin units to which the remaining spin units are related. This prompts the question of the extent to which this offers a solution to the quantum measurement problem, which we leave for future research.


In order to fully reproduce quantum mechanics, accounting for preparations and measurements would not be enough. What about the other fundamental evolution in ``textbook quantum mechanics'', namely unitary evolution? Although the original proposal of spin networks does not seem close to the possibility of reproducing an unitary evolution, we can find clear connections between spin networks and the recent results in \citep{Rudolph_2005}. There, it is shown that any unitary operation over a finite dimensional Hilbert space~$H$ can be reproduced, to arbitrary precision, with the use exclusively of a sequence of \emph{postselected} singlet/triplet measurements over pairs of qubits corresponding to partitions of a larger (but still finite dimensional) Hilbert space~$H'$ that includes the original space~$H \in H'$, given an adequate initial state in~$H'$. Since any sequence of singlet/triplet measurements can be also reproduced as a trivial quantum angular momentum recoupling, we can conclude that one can also reproduce the effect of any unitary quantum gate to arbitrary precision with the use of a large enough spin network (a spin network can be easily formulated as a linear transformation between finite dimensional Hilbert spaces \citep{Major_1999}).

\begin{figure}[h]
\begin{center}
\includegraphics[width=0.6\linewidth]{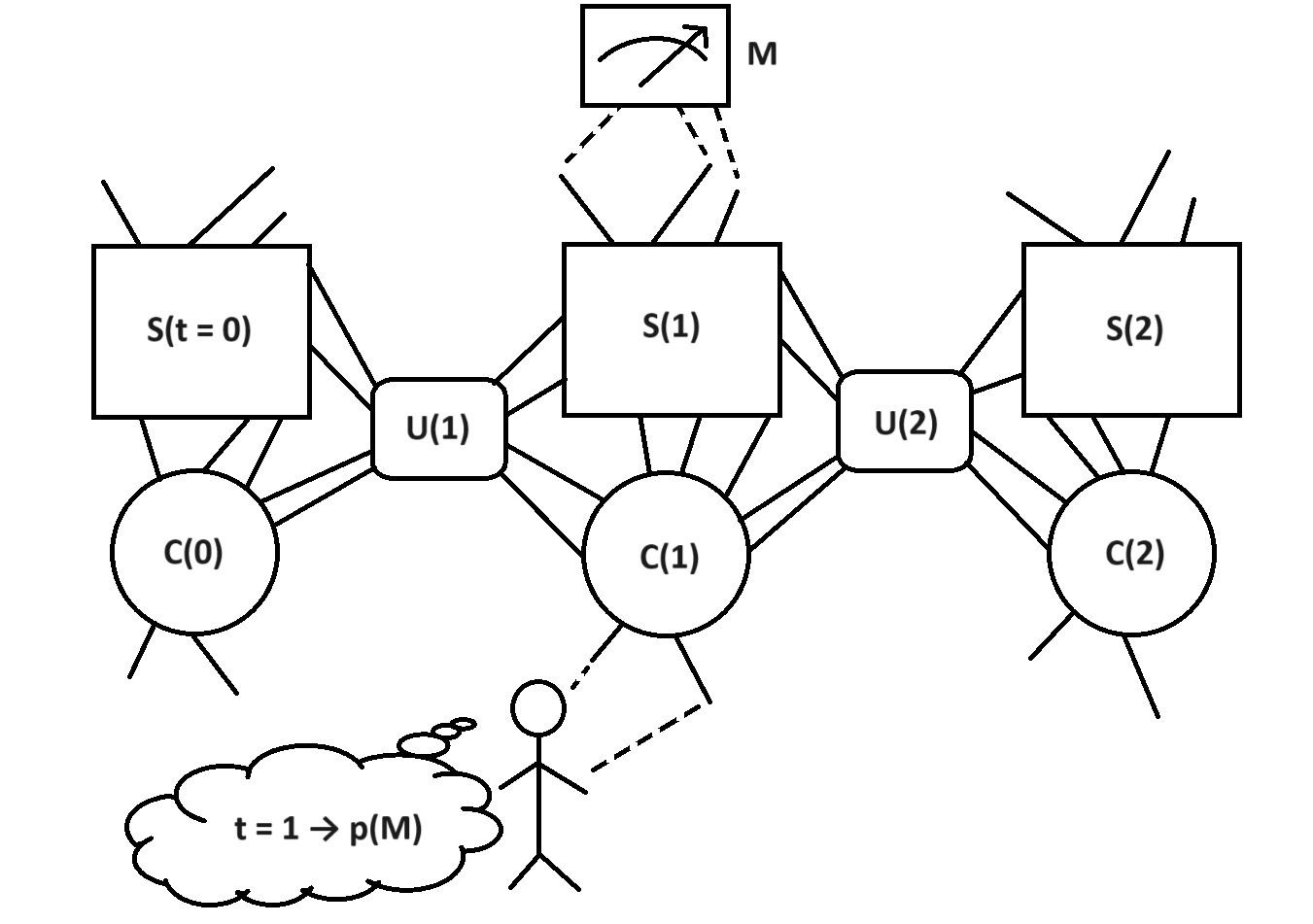}
\caption{A schematic depiction of a \emph{tentative} form of ``discretized approximate'' unitary evolution in time, build up with spin networks. The areas denoted $S(t)$ and $C(t)$ represent spin networks that account for the system (object-context entity) and the clock, respectively, at different steps, and $U(t)$ represents a spin network evolving the system and the clock to the next step. An external agent could directly implement some measurement over the system with a measurement apparatus, obtaining some outcome~$M$. This would however in general alter the system and interrupt the unitary evolution. But the agent can also access the clock and read only the time without significantly altering the system. This information she can use, given that she knows the dynamics, to compute the probabilities $p(M)$ of different possible measurements and outcomes, which characterize the state of the system.}
\label{unitary}
\end{center}
\vspace{-0.5cm}
\end{figure}

It is not trivial though to visualize how an unitary evolution \emph{in time} could be assimilated to an ``evolving'' spin network, that is, a spin network in which different splits and combinations take place sequentially. We shall try nonetheless to give a \emph{tentative} picture on how this could look like. Consider a given initial spin network. The units of a part of it, that we shall call the system~S, will correspond to the inseparable object-context entity. These start to sequentially undergo processes of splits and combinations, enlarging the total spin network, that reproduce certain unitaries, as discussed previously. Now, let us consider that the part of the spin network that undergoes these internal processes is more complicated, and together with the system~S we are interested in, contains something that can be called a clock~C. The idea resembles the Page-Wootters mechanism \citep{PhysRevD.27.2885}: a) C undergoes processes in an specifically intertwined way, and in this sense in a synchronised way, with~S; b) C has a property, called time~$t$ (given for example by the relative angle between two large ends, acting as the hands of a clock) that grows uniformly, at least for a long enough sequence of processes; and c) this~$t$ is such that some external system, for example a physicist, can interact with~C so as to read~$t$ without significantly disturbing the rest of the undergoing processes of~C or~S. Then, the physicist, if she happens to know the dynamics of~S, could compute the probabilities of different results of a measurement that one could force within~S, given a reading of the time~$t$. These time-dependent probabilities could, in some macroscopic limit for~S and~C, reproduce an unitary evolution in time for~S with arbitrary precision. We illustrate this evolution schematically in Fig.~\ref{unitary}.

We shall make an important remark: Unlike the case of the measurement postulate, the unitary evolution in time of textbook quantum mechanics, due to its presentation in the form of a differential equation, makes an unavoidable use of the continuum \emph{in each and every case} where it is applied. Therefore, \emph{any} reproduction of unitary evolution in time with finite spin networks must be \emph{necessarily} only approximate, even if with arbitrary accuracy. In the strict sense, this implies a departure from textbook quantum mechanics, although may not be distinguishable in the experimental range of application of quantum mechanics. Also, a departure from the formal unitary evolution of quantum mechanics does not necessarily mean a fundamentally irreversible evolution. Whether an evolution based on spin networks may destroy and create information at the fundamental level, or may rather have its own form of an information-preserving evolution, different from quantum mechanical unitarity under finite resources, is an interesting question that  goes beyond the discussion presented here.

\section{Conclusions and outlook}

All our statements and the language we use to communicate in everyday life contain implicit or explicit assumptions about infinite resources. When we say that we will meet someone at the Schottentor tram stop in Vienna at 2 p.m., we assume that the location in space-time is given with sufficiently large, practically infinite, resources. The statements we make about quantum systems are no different. When we say ``the spin of a spin one-half particle is oriented `up' along the $x$ direction'', we are expressing specific laboratory arrangements and records in which there are virtually infinite resources for defining position and orientation in the three-dimensional space. These are the orientations of macroscopic devices, such as the Stern-Gerlach magnets and detectors capable of producing detection events (``clicks''). 

We have argued here that ``textbook quantum mechanics'' describes quantum systems in the presence of practically infinite resources, as is the case under standard laboratory conditions. Can one go beyond this? Relational quantum mechanics takes up the challenging task of making sense of quantum theory outside the laboratory, and while it adopts the relational perspective, it applies textbook quantum mechanics to systems with only finite resources, insufficient to make sense of statements like ``spin B is oriented `up' along the direction $x$ relative to spin A''. When A and B are two quantum systems of insufficient dimension there is no resources to define a spatial direction.

We propose here to keep the relational perspective but go a different way in taking the finiteness of resources as an integral part of quantum theory. We give an example of such an approach in terms of the original spin network construction of Penrose. The spin network is a framework in which one expects textbook quantum mechanics and space-time theory to emerge together from more primitive combinatorial rules and only in the limit of infinite resources (infinitely large spin units). Instead of statements like ``the spin is `up' along direction $x$'', we can only make relative statements of the type ``the two spins are aligned'' or ``the two spins are opposite'', with the fundamental impossibility of specifying any spatial direction along which they are aligned, for the simple reason that in general there are no sufficient resources to define the directions. In contrast to RQM, these statements would not be made with respect to any presupposed basis or, equivalently, direction in space (since there are no resources to define one) nor to all of them simultaneously (since this would entail assuming hidden variables). Rather, these statements are made with respect to a rotationally invariant measurement of the total angular momentum of the two spins, with no reference to an external frame of reference.

It has been argued that the original construction of spin networks is formally capable of reproducing both the probabilistic assignments for measurement results and the unitarity of quantum mechanics, provided that sufficiently large networks are employed. From this perspective, textbook quantum mechanics can be seen as the limit of the spin network framework with infinite resources, which is the domain in which quantum pyhsics has been experimentally verified. The question of whether the spin-network construction has physical meaning under strictly finite resources, and whether it can be used to derive potential empirical consequences, is beyond the scope of this conceptual discussion. The aim of this discussion was to use spin networks to shed light on possible alternative approaches to fundamental problems in quantum mechanics, such as contextuality, objectivity or relationalism.

In conclusion, quantum mechanical formalism inherently assumes a macroscopic context, with infinite resources, in the description of any purportedly closed quantum system. These systems would not be in true isolation but in physical relation with the given context. We believe that, together with spin networks formalism, works on \textit{quantum reference frames} \citep{Bartlett_2007,Giacomini_2019,Loveridge_2018,Vanrietvelde_2020}, especially those that are not ``ideal'' and hence does not presuppose infinite resources, may provide further clues as to how to achieve a relational formalization of quantum theory in the presence of finite resources. What would a formalisation of the system-context entity look like, and how would the epistemic status of the entity be specified formally and in language? It is likely that this formalization will move us further away from the visualization so common in classical physics, which implicitly assumes infinite resources and is still somewhat present in textbook quantum mechanics through specific representations of a quantum state (as exemplified through statements such as ``spin is `up' along the direction $\vec{n}$), in much the same way that common language may lose its visualization if the notions that require virtually unlimited resources, such as ``space'', ``time'', ``local'', ``orientation'' or ``distance'' are omitted from the language. Whether this would require a revision of Bohr's dictum from the beginning of the paper remains to be seen.

\section{Acknowledgements}

The authors would like to thank Esteban Castro-Ruiz and Nathan Cohen for their invaluable input during the elaboration of this chapter. Additionally, they would like to thank Flavio Del Santo for his thorough reading and constructive feedback on the initial version of the text. This research was funded in whole or in part by the Austrian Science Fund (FWF) 10.55776/COE1 (Quantum Science Austria), [10.55776/F71] (BeyondC) and [10.55776/RG3] (Reseacrh Group 3). For Open Access purposes, the authors have applied a CC BY public copyright license to any author accepted manuscript version arising from this submission. This publication was made possible through the financial support of the ID 62312 grant from the John Templeton Foundation, as part of The Quantum Information Structure of Spacetime (QISS) Project (qiss.fr). The opinions expressed in this publication are those of the authors and do not necessarily reflect the views of the John Templeton Foundation.


\bibliographystyle{apa-good}
\bibliography{biblio}{}

\end{document}